\documentclass[
aps,%
final,%
notitlepage,%
oneside,%
twocolumn,%
nobibnotes,%
nofootinbib,%
superscriptaddress,%
showpacs,%
centertags,%
amsmath,%
amssymb,%
floatfix]%
{revtex4}

\usepackage{graphicx}
\usepackage{bm}

\newcommand{\PM}{\raisebox{2.5pt}{$+$}\hspace{-2ex}%
\raisebox{-2.5pt}{{\scriptsize(}$\!-\!$\scriptsize{)}}}
\newcommand{\MP}{\raisebox{2.5pt}{$-$}\hspace{-2ex}%
\raisebox{-2.5pt}{{\scriptsize(}$\!+\!$\scriptsize{)}}}
\newcommand{\vect}[1]{\mathbf{#1}}
\newcommand{\ten}[1]{\bm{#1}}
\newcommand{\sprod}{\!\cdot\!}

\newcommand{\vprod}{\!\times\!}
\newcommand{\trace}{\mathrm{Tr}}

\newcommand{\dif}{\mathrm{d}}
\newcommand{\mi}{i}
\newcommand{\me}{e}

\begin{document}

\title{Two-atom van der Waals interaction between
polarizable/magnetizable atoms near magneto-electric bodies}

\author{\firstname{S.~Y.}~\surname{Buhmann}}
\email{s.buhmann@tpi.uni-jena.de}
\affiliation{%
Theoretisch-Physikalisches Institut,
Friedrich-Schiller-Universit\"{a}t Jena,
Max-Wien-Platz 1, D-07743 Jena, Germany
}%
\author{\firstname{H.}~\surname{Safari}}
\affiliation{%
Theoretisch-Physikalisches Institut,
Friedrich-Schiller-Universit\"{a}t Jena,
Max-Wien-Platz 1, D-07743 Jena, Germany
}%
\author{\firstname{Ho Trung}~\surname{Dung}}
\affiliation{%
Institute of Physics, Academy of Sciences and Technology, 1 Mac Dinh
Chi Street, District 1, Ho Chi Minh city, Vietnam
}%
\author{\firstname{D.-G.}~\surname{Welsch}}
\affiliation{%
Theoretisch-Physikalisches Institut,
Friedrich-Schiller-Universit\"{a}t Jena,
Max-Wien-Platz 1, D-07743 Jena, Germany
}%

\begin{abstract}
The van der Waals potential of two atoms in the presence of an
arbitrary arrangement of dispersing and absorbing magneto-electric
bodies is studied. Starting from a polarizable atom placed within a
given geometry, its interaction with a second polarizable/magnetizable
atom is deduced from its Casimir-Polder interaction with a weakly
polarizable/magnetizable test body. The general expressions for the
van der Waals potential hence obtained are illustrated by considering
first the case of two atoms in free space, with special emphasis on
the interaction between (i) two polarizable atoms and (ii) a
polarizable and a magnetizable atom. Furthermore, the influence of
magneto-electric bodies on the van der Waals interaction is studied
in detail for the example of two atoms placed near a perfectly
reflecting plate or a magneto-electric half space, respectively.
\end{abstract}

\pacs{
12.20.-m, %Quantum electrodynamics
42.50.Vk, %Mechanical effects of light on atoms, molecules, electrons,
          %and ions
34.20.-b, %Interatomic and intermolecular potentials and forces,
          %potential energy surfaces for collisions
42.50.Nn  %Quantum optical phenomena in absorbing, dispersive and
          %conducting media
}

\maketitle

%%%%%%%%%%%%%%%%%%%%%%%%%%%%%%%%%%%%%%%%%%%%%%%%%%%%%%%%%%%%%%%%%%%%%%

\section{Introduction}
\label{Sec1}

The van der Waals (vdW) interaction of two neutral, unpolarized, but
polarizable atoms is a well-known consequence of quantum ground-state
fluctuations. For sufficiently small separations, its physical origin
may be seen in the electrostatic Coulomb interaction of the atoms'
fluctuating dipole moments. The vdW interaction was first calculated
in this nonretarded limit by London on the basis of perturbation
theory, who found an attractive potential proportional to $r^{-6}$,
where $r$ denotes the interatomic separation \cite{lon}. For larger
separations, the vacuum fluctuations of the (transverse)
electromagnetic field also contribute to the interaction. This was
first taken into account by Casimir and Polder by means of a
normal-mode expansion of the electromagnetic field, generalizing the
London potential to arbitrary distances and showing that in particular
in the retarded limit the potential varies as $r^{-7}$ \cite{c-p}.

The theory has since been extended with many respects, and various
factors affecting the vdW interaction have been studied. It has been
shown that in the case of one \cite{pass1} or both atoms
\cite{p-t3,shr} being excited, the vdW potential varies as $r^{-6}$
and $r^{-2}$ in the nonretarded and retarded limits, respectively.
Thermal photons present for finite temperature have been found to lead
to a change of the retarded vdW potential of two ground-state atoms
from a $r^{-7}$- to a $r^{-6}$-dependence as soon as the interatomic
separation exceeds the wavelength of the dominant photons
\cite{nin,wen,gdk,bar}. The influence of external electric fields on
the vdW interaction was addressed, where it has been found that the
resulting potential varies as $r^{-3}$ in the nonretarded limit when
the applied field is unidirectional \cite{mil}. Generalizations of the
vdW interaction to the three- \cite{Axilrod43,Aub60,Cirone96,pass2}
and $N$-atom case \cite{p-t1,p-t2} were studied first in the
nonretarded limit and later for arbitrary interatomic separations,
where the potentials were seen to depend on the relative positions of
the atoms in a rather complicated way.

The two-atom vdW interaction may be strongly affected by the presence
of magneto-electric bodies. This was first demonstrated by Mahanty
and Ninham, who employed a semiclassical approach to obtain a general
expression for the vdW potential of two ground-state atoms in the
presence of electric bodies \cite{Mahanty72,mah,mah1976}, and
applied it to the case of two atoms placed between two perfectly
conducting plates \cite{mah}. The situation of two atoms between two
perfectly conducting plates was later reconsidered taking into account
finite temperature effects \cite{bos}. Other scenarios such as two
atoms placed within a planar electric three-layer geometry
\cite{mar} or two anisotropic molecules in front of an electric half
space or within a planar electric cavity have also been studied
\cite{Cho}.

Bearing in mind that the vdW potential of two polarizable atoms may be
modified due to finite temperature, external fields, or the presence
of electric bodies, but remains attractive in all of these cases, it
is rather surprising that the interaction of a polarizable atom with a
magnetizable one is repulsive. This was first realized by Feinberg and
Sucher who restricted their attention to the retarded case and found a
repulsive potential proportional to $r^{-7}$ \cite{feinberg}. Their
result was later extended to all distances \cite{Boyer69,Feinberg70},
and in particular it was shown that the nonretarded potential is
proportional to $r^{-4}$ \cite{Farina02}.

Van der Waals interactions of two atoms exhibiting electric as well
as magnetic properties have so far only been studied in the
free-space case. A much richer range of phenomena is to be expected
when allowing for the presence of magneto-electric bodies, where a
complex interplay of the electric and magnetic properties of the
atoms and the bodies influences both sign and functional dependence
of the two-atom vdW potential. This problem is addressed in the
current work, where we derive the vdW potential of a polarizable atom
with another polarizable or magnetizable one by starting from its
Casimir-Polder (CP) interaction with a weakly polarizable or
magnetizable body, respectively. This approach, which has the
advantage of being much simpler than perturbative methods and
easily applicable to magnetizable atoms, renders general expressions
for the two-atom vdW potentials of polarizable and/or magnetizable
atoms in the presence of an arbitrary arrangement of magneto-electric
bodies, as is shown in Sec.~\ref{Sec2}. In Sec.~\ref{Sec3}, the
general results are applied to the vdW interaction between
polarizable/magnetizable atoms in free space and in front of a
magneto-electric plate. A summary is given in Sec.~\ref{Sec4}.

%%%%%%%%%%%%%%%%%%%%%%%%%%%%%%%%%%%%%%%%%%%%%%%%%%%%%%%%%%%%%%%%%%%%%%

\section{General theory}
\label{Sec2}

Consider first a neutral, nonpolar, ground-state atom or molecule $A$
(briefly referred to as atom in the following) in the presence of
an arbitrary arrangement of dispersing and absorbing magneto-electric
bodies. The atom is characterized by its center-of-mass position
$\mathbf{r}_A$ and its frequency-dependent electric polarizability
$\alpha_A(\omega)$, while the bodies are given by their macroscopic
(relative) permittivity $\varepsilon(\mathbf{r},\omega)$ and
permeability $\mu(\mathbf{r},\omega)$, which are spatially varying,
complex-valued functions of frequency, with the corresponding
Kramers-Kronig relations being satisfied.

Due to the presence of the bodies, the atom will be subject to a CP
force $\mathbf{F}_A$. Within the framework of macroscopic QED in
linear, causal media and by using leading-order perturbation theory,
it can be shown that this force follows from the associated potential
\cite{Buhmann04, Buhmann05}
\begin{equation}
\label{Eq1}
U_A(\mathbf{r}_\mathrm{A})=\frac{\hbar\mu_0}{2\pi}
 \int_0^{\infty}\mathrm{d} u \,u^2 \alpha_A(iu)
 \,\mathrm{Tr}\,
 \bm{G}^{(1)}(\mathbf{r}_A,\mathbf{r}_A,iu)
\end{equation}
according to
\begin{equation}
\label{Eq2}
\mathbf{F}_A
 =-\bm{\nabla}_{\!A}U_A(\mathbf{r}_A)
\end{equation}
($\bm{\nabla}_{\!A}$ $\!\equiv$ $\!\bm{\nabla}_{\!\mathbf{r}_A}$). In
Eq.~(\ref{Eq1}), $\bm{G}^{(1)}(\mathbf{r},\mathbf{r}',\omega)$ is the
scattering part of the classical Green tensor of the electromagnetic
field,
\begin{equation}
\label{Eq3}
\bm{G}(\mathbf{r},\mathbf{r}',\omega)
=\bm{G}^{(0)}(\mathbf{r},\mathbf{r}',\omega)
+\bm{G}^{(1)}(\mathbf{r},\mathbf{r}',\omega)
\end{equation}
[$\bm{G}^{(0)}(\mathbf{r},\mathbf{r}',\omega)$, bulk part], which is
the solution to the equation
\begin{equation}
\label{Eq4}
\biggl[\bm{\nabla}\times
\kappa(\mathbf{r},\omega)
\bm{\nabla}\times\,
 -\frac{\omega^2}{c^2}\,\varepsilon(\mathbf{r},\omega)\biggr]
 \bm{G}(\mathbf{r},\mathbf{r}',\omega)
 =\bm{\delta}(\mathbf{r}-\mathbf{r}')
\end{equation}
[$\kappa(\mathbf{r},\omega)$ $\!=$ $\!\mu^{-1}(\mathbf{r},\omega)$]
together with the boundary condition
\begin{equation}
\label{Eq5}
\bm{G}(\mathbf{r},\mathbf{r}',\omega)\to 0
\quad\mbox{for}\quad|\mathbf{r}-\mathbf{r}'|\to\infty.
\end{equation}

In order to derive the vdW interaction of atom $A$ with a second
polarizable atom $B$, we now introduce an additional, weakly
polarizable body of (small) volume $V_\mathrm{b}$, consisting of a
collection of atoms of type $B$. Provided that the atomic number
density $\eta_B(\vect{r})$ is sufficiently small, the electric
susceptibility of the additional body can be approximated by
\begin{equation}
\label{Eq6}
\chi_\mathrm{e}(\mathbf{r},\omega)
 =\begin{cases}
 \varepsilon_0^{-1}
\eta_B(\vect{r})
 \alpha_B(\omega)&\mbox{if }\mathbf{r}\in V_\mathrm{b},\\
 0&\mbox{if }\mathbf{r}\not\in V_\mathrm{b},
\end{cases}
\end{equation}
so that the permittivity of the total arrangement of bodies reads
$\varepsilon(\mathbf{r},\omega)$ $\!+$
$\!\chi_\mathrm{e}(\mathbf{r},\omega)$, and the corresponding Green
tensor is given by Eq.~(\ref{Eq4}) with
$\varepsilon(\mathbf{r},\omega)$ $\!+$
$\!\chi_\mathrm{e}(\mathbf{r},\omega)$ instead of
$\varepsilon(\mathbf{r},\omega)$. A linear expansion of this
differential equation in terms of $\chi_\mathrm{e}$ reveals that the
presence of the additional body leads to a change of the Green tensor,
whose leading term is
\begin{align}
\label{Eq7}
&\Delta\bm{G}(\mathbf{r},\mathbf{r}',\omega)\nonumber\\
&\quad=\Bigl(\frac{\omega}{c}\Bigr)^2
 \int\mathrm{d}^3s\,
 \chi_\mathrm{e}(\mathbf{s},\omega)
 \bm{G}(\mathbf{r},\mathbf{s},\omega)\!\cdot\!
 \bm{G}(\mathbf{s},\mathbf{r}',\omega).
\end{align}
According to Eq.~(\ref{Eq1}), the resulting change of the CP potential
is
\begin{align}
\label{Eq8}
\Delta U_A(\vect{r}_A)
 =&-\frac{\hbar}{2\pi\varepsilon_0}
 \int_0^\infty\dif u\,\Bigl(\frac{u}{c}\Bigr)^4\alpha_A(\mi u)
 \int\dif^3s\,\chi_\mathrm{e}(\vect{s},\mi u)
 \nonumber\\
&\times\trace\bigl[
 \ten{G}(\vect{r}_A,\vect{s},\mi u)\sprod
 \ten{G}(\vect{s},\vect{r}_A,\mi u)
 \bigr].
\end{align}
Recalling Eq.~(\ref{Eq6}), on can easily see that $\Delta U_A$ is just
an integral over two-atom potentials $U_{AB}(\vect{r}_A,\vect{r}_B)$,
\begin{equation}
\label{Eq9}
\Delta U_A(\vect{r}_A)
 =\int_{V_\mathrm{b}}\dif^3r_B\,\eta_B(\vect{r}_B)
U_{AB}(\vect{r}_A,\vect{r}_B),
\end{equation}
where
\begin{multline}
\label{Eq10}
U_{AB}(\mathbf{r}_A,\mathbf{r}_B)
 =-\frac{\hbar}{2\pi\varepsilon_0^2}\int_0^\infty\mathrm{d}u\,
 \Bigl(\frac{u}{c}\Bigr)^4\alpha_{A}(iu)\alpha_{B}(iu)\\
 \times\mathrm{Tr}\bigl[
 \bm{G}(\mathbf{r}_A,\mathbf{r}_B,iu)
 \!\cdot\!\bm{G}(\mathbf{r}_B,\mathbf{r}_A,iu)
 \bigr].
\end{multline}
is the vdW potential between two polarizable atoms in the presence of
an arbitrary arrangement of dispersing and absorbing magneto-electric
bodies. The total force acting on atom $A$($B$) due to atom $B$($A$)
and the bodies is just the sum of the single-atom CP force (\ref{Eq2})
and the two-atom vdW force
\begin{equation}
\label{inforce}
\mathbf F_{AB(BA)}
=-\bm \nabla_{\!A(B)}U_{AB}(\mathbf{r}_A,\mathbf{r}_B),
\end{equation}
where in general $\mathbf F_{AB}$ $\!\neq$ $-\mathbf F_{BA}$, due to
the presence of the bodies. Equation~(\ref{Eq10}) agrees with the
result obtained on the basis of fourth-order perturbation theory, with
the perturbative calculation being much more lengthy
\cite{Buhmann06,Safari06}. Needless to say that the method presented
here can be easily extended to derive $N$-atom potentials
\cite{Buhmann06,Buhmann06b}, whereas the perturbative method becomes
increasingly cumbersome for large $N$.

The vdW potential between a polarizable atom $A$ and a magnetizable
atom $B$ can be derived in a very analogous way. Again we start from
atom $A$ placed within an arbitrary arrangement of magneto-electric
bodies, Eqs.~(\ref{Eq1}) and~(\ref{Eq4}), but this time we add a
weakly magnetizable body, consisting of a collection of magnetizable
atoms of type $B$. For sufficiently small atomic number density the
magnetic susceptibility of this body approximately reads
\begin{equation}
\label{Eq13}
\chi_\mathrm{m}(\mathbf{r},\omega)
 =\begin{cases}
 \mu_0\eta_B(\vect{r})
 \beta_B(\omega)&\mbox{if }\mathbf{r}\in V_\mathrm{b},\\
 0&\mbox{if }\mathbf{r}\not\in V_\mathrm{b},
\end{cases}
\end{equation}
with $\beta_B(\omega)$ denoting the magnetizability. The inverse
permeability of the total arrangment of bodies reads
$\kappa(\mathbf{r},\omega)$ $\!-$
$\!\chi_\mathrm{m}(\mathbf{r},\omega)$, so that the Green tensor
corresponding to this arrangement is given by Eq.~(\ref{Eq4})
with \mbox{$\kappa(\mathbf{r},\omega)$ $\!-$
$\!\chi_\mathrm{m}(\mathbf{r},\omega)$} instead of
$\kappa(\mathbf{r},\omega)$. A linear expansion in terms of
$\chi_\mathrm{m}(\mathbf{r},\omega)$ leads to
\begin{multline}
\label{Eq14}
\Delta\bm{G}(\mathbf{r},\mathbf{r}',\omega)=\\
-\int\mathrm{d}^3s\,
 \chi_\mathrm{m}(\mathbf{s},\omega)
 \bigl[\bm{G}(\mathbf{r},\mathbf{s},\omega)\vprod
 \overleftarrow{\bm{\nabla}}_{\!\vect{s}}\bigr]
 \cdot\!\bm{\nabla}_{\!\vect{s}}\vprod
 \bm{G}(\mathbf{s},\mathbf{r}',\omega).
\end{multline}
Combining this with Eq.~(\ref{Eq1}), we find that the change in the CP
potential due to the presence of the additional magnetizable body is
\begin{align}
\label{Eq15}
&\Delta U_A(\vect{r}_A) =
-\frac{\hbar}{2\pi\varepsilon_0}
 \int_0^\infty\!\!\dif u\,\Bigl(\frac{u}{c}\Bigr)^2\alpha_A(\mi u)
 \int\!\dif^3s\,\chi_\mathrm{m}(\vect{s},\mi u)\nonumber\\
&\quad\times\trace\Bigl\{
 \bigl[\ten{G}(\vect{r}_A,\vect{s},\mi u)
 \vprod\overleftarrow{\bm{\nabla}}_{\!\vect{s}}\bigr]
 \sprod\bm{\nabla}_{\!\vect{s}}\vprod
 \ten{G}(\vect{s},\vect{r}_A,\mi u)\Bigr\}.
\end{align}
Finally, upon using Eq.~(\ref{Eq13}), Eq.~(\ref{Eq15}) can be
rewritten in the form of Eq.~(\ref{Eq9}), where now
\begin{align}
\label{Eq16}
&U_{AB}(\vect{r}_A,\vect{r}_B)
 =-\frac{\hbar\mu_0}{2\pi\varepsilon_0}
 \int_0^\infty\dif u\,\Bigl(\frac{u}{c}\Bigr)^2
 \alpha_A(\mi u)
 \beta_B(\mi u)\nonumber\\
&\quad\times\trace\bigl[
 \ten{G}(\vect{r}_A,\vect{r},\mi u)
 \vprod\overleftarrow{\bm{\nabla}}_{\!\vect{r}}
 \sprod\bm{\nabla}_{\!\vect{r}}\vprod
 \ten{G} (\vect{r},\vect{r}_A,\mi u)
 \bigr]_{\vect{r}=\vect{r}_B}
%.
\end{align}
is the vdW potential between a polarizable atom $A$ and a magnetizable
atom $B$ in the presence of an arbitrary arrangement of
magneto-electric bodies. To our knowledge, Eq.~(\ref{Eq16}) has never
been derived so far.

%%%%%%%%%%%%%%%%%%%%%%%%%%%%%%%%%%%%%%%%%%%%%%%%%%%%%%%%%%%%%%%%%%%%%%

\section{Examples}
\label{Sec3}

%%%%%%%%%%%%%%%%%%%%%%%%%%%%%%%%%%%%%%%%%%%%%%%%%%%%%%%%%%%%%%%%%%%%%%

\subsection{Free space}
\label{Sec3.1}

In order to illustrate the two-atom vdW potentials (\ref{Eq10}) and
(\ref{Eq16}), let us first consider two atoms in free space, with the
Green tensor being given by \cite{Knoll01}
\begin{gather}
\label{Eq17}
\ten{G}^{(0)}(\vect{r},\vect{r}',\mi u)
=\frac{\me^{-u\rho/c}}{4\pi\rho}
 \bigl\{a[c/(u\rho)]\ten{I}
 -b[c/(u\rho)]
 \vect{e}_\rho\vect{e}_\rho\bigr\},\\
\label{Eq18}
a(x)=1+x+x^2,\qquad
b(x)=1+3x+3x^2
\end{gather}
($\bm{\rho}$ $\!\equiv$ $\!\vect{r}-\vect{r}'$; $\rho$ $\!\equiv$
$\!|\bm{\rho}|$; $\vect{e}_\rho$ $\!\equiv$ $\!\bm{\rho}/\rho$;
$\ten{I}$, unit tensor). In this case, the vdW potential between
two polarizable atoms, Eq.~(\ref{Eq10}), reads
\begin{gather}
\label{Eq19}
U^{(0)}_{AB}(\vect{r}_A,\vect{r}_B)
 =-\frac{\hbar}{32\pi^3\varepsilon_0^2l^6}
 \int_0^\infty\!\dif u\,
 \alpha_A(\mi u)\alpha_B(\mi u)g(ul/c),\\
\label{Eq20}
g(x)=2e^{-2x}(3+6x+5x^2+2x^3+x^4)
\end{gather}
($l$ $\!\equiv$ $|\mathbf{r}_A-\mathbf{r}_B|$), in agreement with the
well-known result of Casimir and Polder \cite{c-p}. From
\begin{align}
\label{Eq21}
\bm{\nabla}\biggl[\frac{g(ur/c)}{r^6}\biggr]
=&-\frac{4\hat{\vect{r}}}{r^7}
 \bigl[e^{-2x}(9+18x+16x^2\nonumber\\
&+8x^3+3x^4+x^5)\bigr]_{x=ur/c}
\end{align}
it can be seen that the potential~(\ref{Eq19}) is always attractive.

In the retarded limit, where $l$ $\!\gg$ $\!
c/\omega_{\mathrm{min}}$ ($\omega_{\mathrm{min}}$ denoting the minimum
of all resonance frequencies of atoms $A$ and $B$) the exponential
factor effectively limits the $u$-integral in Eq.~(\ref{Eq19}) to a
region where
\begin{equation}
\label{alpha0}
\alpha_{A}(iu)\simeq\alpha_{A}(0),\quad
\alpha_{B}(iu)\simeq\alpha_{B}(0),
\end{equation}
hence the integral can be performed in closed form to yield
\begin{gather}
\label{E35}
U^{(0)}_{AB}(\vect{r}_A,\vect{r}_B)=-\frac{C_7}{l^7}\,,\\[1ex]
\label{cr}
C_7=\frac{23\hbar c\alpha_{A}(0)\alpha_{B}(0)}
{64\pi^3\varepsilon_0^2}\,.
\end{gather}
In the opposite, nonretarded limit, where $l$ $\!\ll$
$\!c/\omega_\mathrm{max}$ ($\omega_{\mathrm{max}}$ denoting the
maximum of all resonance frequencies of atoms $A$ and $B$), the
factors $\alpha_A(\mi u)$ and $\alpha_B(\mi u)$ effectively limit
the $u$-integral in Eq.~(\ref{Eq19}) to a region where
\begin{equation}
\label{Eq25}
g(ul/c)\simeq g(0)=6,
\end{equation}
so that the London potential \cite{lon} is recovered,
\begin{gather}
\label{E36}
U^{(0)}_{AB}(\vect{r}_A,\vect{r}_B)=-\frac{C_6}{l^6}\,,\\[1ex]
\label{cnr}
C_6=\frac{3\hbar}{16\pi^3\varepsilon_0^2}
 \int_0^\infty {\rm d}u\,\alpha_{A}(iu)\alpha_{B}(iu).
\end{gather}

To calculate the potential between a polarizable atom and a
magnetizable one, we first use Eq.~(\ref{Eq17}) to derive
\begin{gather}
\label{Eq28}
\bm{\nabla}\vprod
 \ten{G}^{(0)}(\vect{r},\vect{r}',\mi u)
 =-\frac{\me^{-\frac{u\rho}{c}}}{4\pi\rho^2}
 \Bigl(1+\frac{u\rho}{c}\Bigr)\vect{e}_\rho\vprod\ten{I},\\[.5ex]
\label{Eq29}
\ten{G}^{(0)}(\vect{r},\vect{r}',\mi u)
 \vprod\overleftarrow{\bm{\nabla}}{}'
 =\frac{\me^{-\frac{u\rho}{c}}}{4\pi\rho^2}
 \Bigl(1+\frac{u\rho}{c}\Bigr)\ten{I}\vprod\vect{e}_\rho.
\end{gather}
Substituting Eqs.~(\ref{Eq28}) and (\ref{Eq29}) into
Eq.~(\ref{Eq16}) and using the identity
\begin{equation}
\label{Eq30}
\trace\bigl[\vect{e}_\rho\vprod
 \ten{I}\vprod\vect{e}_\rho\bigr]=-2,
\end{equation}
we find that
\begin{multline}
\label{Eq40}
U_{AB}^{(0)}(\vect{r}_A,\vect{r}_B) =\\
\frac{\hbar\mu_0}{32\pi^3\varepsilon_0l^4}
 \int_0^{\infty}\dif u\,\Bigl(\frac{u}{c}\Bigr)^2
 \alpha_{A}(\mi u)\beta_{B}(\mi u) h(ul/c),
\end{multline}
\begin{equation}
 h(x) = 2e^{-2x}(1+2x+x^2),
\end{equation}
in agreement with results found earlier \cite{Boyer69,Feinberg70}.

In contrast to the attractive vdW potential between two polarizable
atoms, Eq.~(\ref{Eq19}), the vdW potential between a polarizable and a
magnetizable atom, Eq.~(\ref{Eq40}), is always repulsive, as can be
seen from
\begin{equation}
\label{Eq41}
\bm{\nabla}\biggl[\frac{h(ur/c)}{r^4}\biggr]
 =-\frac{4\hat{\vect{r}}}{r^5}
 \bigl[e^{-2x}(2+4x+3x^2+x^3)\bigr]_{x=ur/c}.
\end{equation}
In particular, one finds, on using Eqs.~(\ref{alpha0}) and
\begin{equation}
\label{Eq33}
h(ul/c)\simeq h(0)=2,
\end{equation}
that Eq.~(\ref{Eq40}) reduces to
\begin{gather}
\label{Eq42}
U_{AB}^{(0)}(\vect{r}_A,\vect{r}_B)
 =\frac{C_7}{l^7}\,,\\[1ex]
\label{Eq43}
C_7=\frac{7\hbar c\mu_0\alpha_{A}(0)\beta_{B}(0)}
 {64\pi^3\varepsilon_0}
\end{gather}
and
\begin{gather}
\label{Eq44}
U_{AB}^{(0)}(\vect{r}_A,\vect{r}_B)
=\frac{C_4}{l^4}\,,\\[1ex]
C_4=\frac{\hbar\mu_0}{16\pi^3\varepsilon_0}
 \int_0^\infty\dif u\,\Bigl(\frac{u}{c}\Bigr)^2
 \alpha_{A}(\mi u)\beta_{B}(\mi u)
\end{gather}
in the retarded and nonretarded limits, respectively. The vdW
potential between a polarizable atom and a magnetizable one hence
shows a $l^{-7}$ power law in the nonretarded limit which is weaker
than the corresponding $l^{-7}$ vdW potential of two polarizable atoms
by a factor of $7/23$; while in the nonretarded limit, the potential
between a polarizable and a magnetizable atom follows a $l^{-4}$ power
law which is more weakly diverging than the corresponding $l^{-6}$
potential between two polarizable atoms.

%%%%%%%%%%%%%%%%%%%%%%%%%%%%%%%%%%%%%%%%%%%%%%%%%%%%%%%%%%%%%%%%%%%%%%

\subsection{Semi-infinite half space}
\label{Sec3.2}

Let us now study the influence of the presence of magneto-electric
bodies on the two-atom vdW potential. To that end, we consider two
polarizable atoms placed near a homogeneous semi-infinite half space.
We choose the coordinate system such that the $z$ axis is
perpendicular to the plate with the origin being on its surface, and
the two atoms lie in the $xz$ plane (Fig.~\ref{reflector}).  In this
case, the nonzero elements of the scattering Green tensor are given by
(App.~\ref{mulg})
\begin{align}
\label{62}
&{G}^{(1)}_{xx(yy)}({\mathbf r}_{A},
 {\mathbf r}_{B},iu)=\frac{1}{8\pi}\int_0^\infty {\rm d}q\,q
 e^{-bZ_+}\nonumber\\[.5ex]
&\quad\times\biggl[\frac{J_0(qX)\;\PM\,
 J_2(qX)}{b}\,r_s
 -\frac{b[J_0(qX)
 \;\MP\,J_2(qX)]}{k^2}\,r_p\biggr],\\
\label{63}
&{G}^{(1)}_{xz(zx)}({\mathbf r}_{A},
 {\mathbf r}_{B},iu)=\MP
 \frac{1}{4\pi}\!\int_0^\infty\!\! {\rm d}q\,q^2
 e^{-bZ_+}\frac{J_1(qX)}{k^2}\,r_{p},\\
\label{64}
&{G}^{(1)}_{zz}({\mathbf r}_{A},{\mathbf r}_{B},iu)
 =-\frac{1}{4\pi}\int_0^\infty {\rm d}q\,q^3
 e^{-bZ_+}\frac{J_0(qX)}{bk^2}\,r_p
\end{align}
[$J_\nu(x)$, Bessel function; $Z_+$ $\!=$ $\!z_{A}$ $\!+$
$\!z_{B}$; $X$ $\!=$ $\!x_{B}$ $\!-$ $\!x_{A}$], where $r_\sigma$
$\!=$ $r_\sigma(q,u)$ ($\sigma$ $\!=$ $\!s,p$) are the reflection
coefficients of the half space [cf.~Eqs.~(\ref{rp-ref}), (\ref{eq95})
and (\ref{eq95b}) below], and
\begin{gather}
\label{bl}
b=b(q,u)=\sqrt{\frac{u^2}{c^2}+q^2}\,,\\[.5ex]
\label{kl}
k=k(q,u)=\frac{u}{c}\,.
\end{gather}

According to the decomposition (\ref{Eq3}) of the Green tensor, the
two-atom potential~(\ref{Eq10}) can be split into three parts,
\begin{align}
\label{uparts}
U_{AB}(\mathbf{r}_A,\mathbf{r}_B)
=&\,U_{AB}^{(0)}(\mathbf{r}_A,\mathbf{r}_B)
 +U_{AB}^{(1)}(\mathbf{r}_A,\mathbf{r}_B)
\nonumber\\[.5ex]
&+U_{AB}^{(2)}(\mathbf{r}_A,\mathbf{r}_B),
\end{align}
where the bulk-part contribution $U_{AB}^{(0)}(\mathbf r_{A},\mathbf
r_{B})$ is simply the free-space result~(\ref{Eq19}),
\begin{align}
\label{81}
&U_{AB}^{(1)}(\mathbf r_{A},\mathbf
r_{B})=-\frac{\hbar}{\pi\varepsilon_0^2}\int_0^\infty {\mathrm d}u\,
\Bigl(\frac{u}{c}\Bigr)^4\alpha_{A}(iu)\alpha_{B}(iu)\nonumber\\
&\quad\times
\mathrm{Tr}\big[{\bm G}^{(0)}(\mathbf r_{A},
 \mathbf r_{B},iu)\!\cdot\!{\bm G}^{(1)}
 (\mathbf r_{B},\mathbf r_{A},iu)\big]\nonumber\\
&=-\frac{\hbar}{32\pi^3\varepsilon_0^2l}\int_0^\infty {\mathrm  d}
u\,\Bigl(\frac{u}{c}\Bigr)^4\alpha_{A}(iu)\alpha_{B}(iu)\,
e^{-ul/c}\nonumber\\
&\quad\times\int_0^\infty {\rm d}q\,q
e^{-bZ_+}
\left(
\bigg\{\bigg[2f(\xi)-g
(\xi)
\frac{X^2}{l^2}\bigg]\bigg[\frac{r_s}{b}-\frac{br_p}{k^2}
\bigg]\right.\nonumber\\
&\quad -2\bigg[f(\xi)-g(\xi)
\frac{Z^2}{l^2}\bigg]
\frac{q^2r_p}{bk^2}\bigg\}J_0(qX)\nonumber\\
&\quad\left.
-g(\xi)\frac{X^2}{l^2}\bigg[\frac{r_s}{b}+
\frac{br_p}{k^2}\bigg]J_2(qX)
\right)
\end{align}
comes from the cross term of bulk and scattering parts [with
$Z$ $\!=$ $\!z_{B}$ $\!-$ $\!z_{A}$, $\xi$ $\!=$ $\!c/(lu)$, and
$a(x)$ and $b(x)$ as given in Eq.~(\ref{Eq18})], and
\begin{align}
\label{82}
&U_{AB}^{(2)}(\mathbf r_{A},\mathbf r_{B})
 =-\frac{\hbar}{2\pi\varepsilon_0^2}\int_0^\infty \!{\mathrm d}u\,
 \Bigl(\frac{u}{c}\Bigr)^4\alpha_{A}(iu)\alpha_{B}(iu)\nonumber\\
&\quad\times
 \mathrm{Tr}\big[{\bm G}^{(1)}(\mathbf r_{A},
 \mathbf r_{B},iu)\!\cdot\!{\bm G}^{(1)}
 (\mathbf r_{B},\mathbf r_{A},iu)\big]\nonumber\\
&=-\frac{\hbar}{64\pi^3\varepsilon_0^2}\int_0^\infty\!\!{\mathrm d}u
 \Bigl(\frac{u}{c}\Bigr)^4\alpha_{A}(iu)\alpha_{\mathrm B}(iu)
 \int_0^\infty \!\!{\rm d}q\, q
 \int_0^\infty \!\!{\rm d}q'\,q' \nonumber\\
&\quad\times
 e^{-(b+b')Z_+}
 \biggl\{
 \bigg[\frac{r_sr'_s}{bb'}
 + \frac{r_pr'_p}{k^4}\bigg(bb'
 +\frac{2q^2q^{\prime 2}}{bb'}\bigg)
-\frac{b'r_sr'_p}{b{k}^2}\nonumber\\
&\quad
-\frac{br_pr'_s }{b'{k}^2}\bigg]
 J_0(qX)J_0(q' X)+\frac{4qq'
 r_pr'_p}{k^4}
 J_1(qX)J_1(q' X)\nonumber\\
&\quad
 + \bigg[\frac{r_sr'_s}{bb'}
 +\frac{bb'r_pr'_p}{k^4}
 +\frac{b'r_sr'_p}{b{k}^2}+\frac{b
 r_pr'_s }{b'{k}^2}\bigg]\!
 J_2(qX)J_2(q' X)\!\biggr\}
\end{align}
is the scattering-part contribution [$b'$ $\!=$ $\!b(q',u)$,
$r'_\sigma$ $\!=r_\sigma(q',u)$ for $\sigma$ $\!=$ $\!s,p$].

%%%%%%%%%%%%%%%%%%%%%%%%%%%%%%%%%%%%%%%%%%%%%%%%%%%%%%%%%%%%%%%%%%%%%%

\subsubsection{Perfectly reflecting plate}
\label{Sec3.2.1}

%%%%%%%%%%%%%%%%%%%%%%%%%%%%%%%% FIGURE PERFECT REFLECTOR %%%%%%%%%%%%
\begin{figure}[!t]
\noindent
\begin{center}
\includegraphics[width=.9\linewidth]{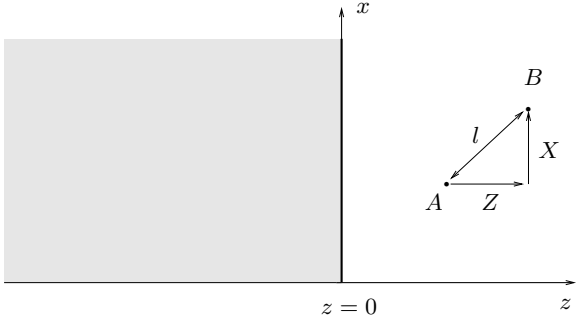}
\end{center}
\caption{
\label{reflector}
Two atoms near a perfectly reflecting plate.}
\end{figure}%
%%%%%%%%%%%%%%%%%%%%%%%%%%%%%%%%%%%%%%%%%%%%%%%%%%%%%%%%%%%%%%%%%%%%%%
For a perfectly reflecting plate, the reflection coefficients are
simply given by
\begin{equation}
\label{rp-ref}
r_s=\mp 1,\qquad r_p=\pm 1,
\end{equation}
where the upper (lower) sign corresponds to a perfectly conducting
(permeable) plate. In the retarded limit, where $l,z_{A},z_{B}$
$\!\gg$ $\!c/\omega_{\mathrm{min}}$, $U_{AB}^{(0)}$ is given by
Eq.~(\ref{E35}), whereas $U_{AB}^{(1)}$ [Eq.~(\ref{81})] and
$U_{AB}^{(2)}$ [Eq.~(\ref{82})] can be given in closed form
only in some special cases. If \mbox{$X$ $\!\ll$ $\!Z_+$}
(cf.~Fig.~\ref{reflector}), we derive, on using the relevant elements
of the scattering Green tensor as given in App.~\ref{mulg}
[Eqs.~(\ref{rx}) and (\ref{rz})],
\begin{align}
\label{gl79}
& U_{AB}^{(1)}=
\pm\frac{32}{23}\,\frac{(X^2+6l^2)C_7}{l^3Z_+(l+Z_+)^5}\,,
\\[.5ex]
\label{gl80}
& U_{AB}^{(2)}=-\frac{C_7}{Z_+^7}\,,
\end{align}
where $C_7$ is given by Eq.~(\ref{cr}). Thus, recalling
Eq.~(\ref{E35}), the two-atom vdW potential~(\ref{uparts}) reads
\begin{equation}
\label{gl801}
U_{AB}=
-\frac{C_7}{l^7}\pm\frac{32}{23}
\frac{(X^2+6l^2)C_7}{l^3Z_+(l+Z_+)^5}-\frac{C_7}
{Z_+^7}\,.
\end{equation}
In particular, if $z_A/z_B$ $\!\ll$ $\!1$, Eqs.~(\ref{gl79}) and
(\ref{gl80}) imply that
\begin{align}
\label{gl81}
& U_{AB}^{(1)}=\mp\frac{6}{23}\,U_{AB}^{(0)}\,,
\\[.5ex]
\label{gl82}
& U_{AB}^{(2)}=U_{AB}^{(0)}\,,
\end{align}
so the presence of the perfectly reflecting plate leads to
an enhancement of the interaction potential,
\begin{equation}
\label{gl83}
U_{AB}=
\begin{cases}
 \displaystyle
 \frac{40}{23}\,U_{AB}^{(0)},\\[2ex]
 \displaystyle
 \frac{52}{23}\,U_{AB}^{(0)}
\end{cases}
\end{equation}
for a perfectly conducting or permeable plate, respectively.

Quite generally, since the bulk part $U_{AB}^{(0)}$ [first term on the
r.h.s.\ of Eq.~(\ref{gl801})] is negative, the interaction potential
is enhanced (reduced) by the plate if the scattering part
\mbox{$U_{AB}^{(1)}\!+\!U_{AB}^{(2)}$} [second and third terms
on the r.h.s.\ of Eq.~(\ref{gl801})] is negative (positive).
In the case of a perfectly conducting plate, it is seen that
especially for \mbox{$Z$ $\!=$ $\!0$}, briefly referred to as the
parallel case, \mbox{$U_{AB}^{(1)}\!+\!U_{AB}^{(2)}$} is
positive, and hence the interaction potential is reduced by the plate,
whereas for $X$ $\!=$ $\!0$, briefly referred to as the vertical case,
$U_{AB}^{(1)}\!+\!U_{AB}^{(2)}$ is positive and the interaction
potential is reduced iff
\begin{equation}
z_B/z_A\lesssim 4.90,
\end{equation}
where, without loss of generality, atom $A$ is assumed to be closer to
the plate than atom $B$. It is apparent from Eq.~(\ref{gl801}) that
for a perfectly permeable plate $U_{AB}^{(1)}\!+\!U_{AB}^{(2)}$ is
always negative, and hence the interaction potential is always
enhanced by the plate.

In the nonretarded limit, where $l,z_{A},z_{B}$ $\!\ll$
$\!c/\omega_{\mathrm{max}}$, $U_{AB}^{(0)}$ is given by
Eq.~(\ref{E36}), and from Eqs.~(\ref{81}) and (\ref{82}) we derive,
on making use of the relevant elements of the scattering Green tensor
as given in App.~\ref{mulg} [Eqs.~(\ref{xx})--(\ref{zz})],
\begin{align}
\label{e1}
&U_{AB}^{(1)}=\pm
\frac{\bigl[4X^4-2Z^2Z_+^2+X^2(Z_+^2+Z^2)\bigr]C_6}
{3l^5l_+^5}\,,\\
\label{e2}
&U_{AB}^{(2)}=-\frac{C_6}{l_+^6}
\end{align}
($l_+$ $\!=$ $\!\sqrt{X^2+Z_+^2}$), where $C_6$ is given by
Eq.~(\ref{cnr}). Hence, the interaction potential (\ref{uparts}),
reads, on recalling Eq.~(\ref{E36}),
\begin{equation}
\label{full}
U_{AB} =-\frac{C_6}{l^6}
\pm \frac{\bigl[4X^4\!-\!2Z^2Z_+^2\!+\!X^2(Z_+^2\!+\!Z^2)\bigr]C_6}
{3l^5l_+^5}-\frac{C_6}{l_+^6}\,.
\end{equation}

Let us again consider the effect of the plate on the interaction
potential for the parallel and vertical cases. In the parallel case,
Eq.~(\ref{full}) takes the form
\begin{equation}
\label{par}
U_{AB}=-\frac{C_6}{l^6}\pm
\frac{(4l^2+Z_+^2)C_6}{3l^3(l^2+Z_+^2)^{\frac{5}{2}}}
-\frac{C_6}{(l^2+Z_+^2)^3}\,.
\end{equation}
which in the on-surface limit $Z_+$ $\!\to$ $\!0$ approaches
\begin{equation}
\label{gl95}
U_{AB}=
\begin{cases}
 \displaystyle\frac{2}{3}\,U_{AB}^{(0)},\\[2ex]
 \displaystyle\frac{10}{3}\,U_{AB}^{(0)}
\end{cases}
\end{equation}
for a perfectly conducting or permeable plate, respectively.
It can easily be seen that the term $U_{AB}^{(1)}$ [second term on the
r.h.s.\ of Eq.~(\ref{par})] dominates the term $U_{AB}^{(2)}$
[third term on the r.h.s.\ of Eq.~(\ref{par})], so
\mbox{$U_{AB}^{(1)}\!+\!U_{AB}^{(2)}$} is positive (negative)
for a perfectly conducting (permeable) plate, and hence the
interaction potential is reduced (enhanced) due to the presence of the
plate.

In the vertical case, from Eq.~(\ref{full}) the interaction potential
is obtained to be
\begin{equation}
\label{ver}
U_{AB}=-\frac{C_6}{l^6}\mp\frac{2C_6}{3Z_+^3l^3}
 -\frac{C_6}{Z_+^6}\,.
\end{equation}
It is obvious that \mbox{$U_{AB}^{(1)}\!+\!U_{AB}^{(2)}$}
[second and third terms on the r.h.s.\ of Eq.~(\ref{ver})] is negative
when the plate is perfectly conducting, thereby enhancing the
interaction potential since $U_{AB}^{(0)}$ [first term in
Eq.~(\ref{ver})] is negative. In the case of a perfectly permeable
plate, $U_{AB}^{(1)}\!+\!U_{AB}^{(2)}$ is positive iff
\begin{equation}
\label{condition}
\frac{z_{B}}{z_{A}}<1+\frac{2}{(3/2)^{1/3}-1}
 \simeq 14.82,
\end{equation}
where atom $A$ is again assumed to be closer to the plate than atom
$B$.

\begin{figure}[t]
\noindent
\begin{center}
\includegraphics[width=.9\linewidth]{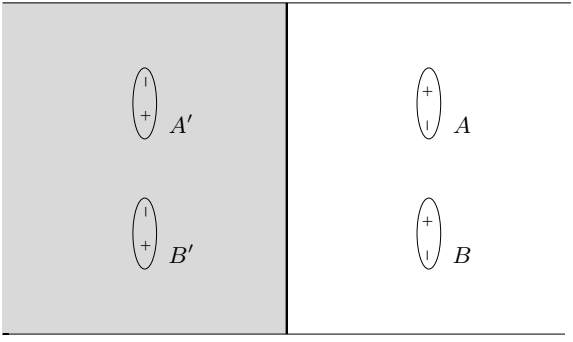}
\end{center}
\caption{
\label{con-par}
Two electric dipoles near a perfectly conducting plate (parallel
case).}
\end{figure}

\begin{figure}[t]
\noindent
\begin{center}
\includegraphics[width=.9\linewidth]{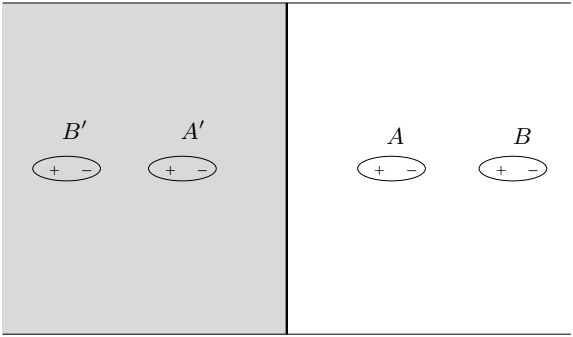}
\end{center}
\caption{
\label{con-ver}
Two electric dipoles near a perfectly conducting plate (vertical
case).}
\end{figure}

\begin{figure}[t]
\noindent
\begin{center}
\includegraphics[width=.9\linewidth]{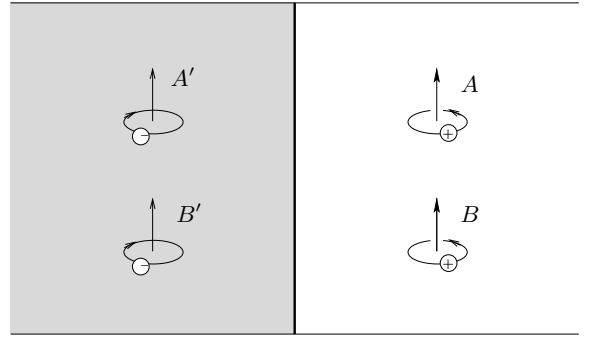}
\end{center}
\caption{
\label{mag-par}
Two magnetic dipoles near a perfectly conducting plate (parallel
case).}
\end{figure}

\begin{figure}[t]
\noindent
\begin{center}
\includegraphics[width=.9\linewidth]{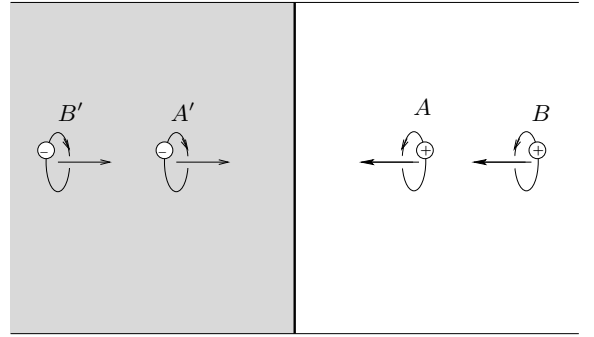}
\end{center}
\caption{
\label{mag-ver}
Two magnetic dipoles near a perfectly conducting plate (vertical
case).}
\end{figure}

The enhancing/reducing effect of the perfectly reflecting plate on
the two-atom vdW potential in the various cases considered can be
systematized in a simple way. Since $U_{AB}^{(0)}$ and $U_{AB}^{(2)}$
are negative in all the cases, the enhancement or reduction of the vdW
potential due to the presence of the plate depends only on the sign of
$U_{AB}^{(1)}$ and its magnitude compared to that of $U_{AB}^{(2)}$.
%%%%%%%%%%%%%%%%%%%%%%%%%%%%%%%%%%%%%%%%%%%%%%%%%%%%%%%%%%%%%%%%%%%%%%
\begin{table}[t]
\begin{tabular}{|c|c|c|}
\hline
{} & conducting plate & permeable plate\\
\hline
parallel case & $+$ & $-$\\
\hline
vertical case & $-$ & $+$\\
\hline
\end{tabular}
\caption{
\label{sign}
Sign of $U_{AB}^{(1)}$ for a perfectly reflecting plate.}
\end{table}%
%%%%%%%%%%%%%%%%%%%%%%%%%%%%%%%%%%%%%%%%%%%%%%%%%%%%%%%%%%%%%%%%%%%%%%
Moreover, the results for the non-retarded limit (the sign of
$U_{AB}^{(1)}$ being summarized in Tab.~\ref{sign}) can be explained
by using the method of image charges, where the two-atom vdW
interaction is regarded as being due to the interactions between
fluctuating dipoles $A$ and $B$ and their images ${A}'$ and ${B}'$ in
the plate, with
\begin{equation}
 \hat H_{\mathrm {int}}=\hat V_{AB}+\hat V_{AB'}+\hat V_{BA'}
\end{equation}
being the corresponding interaction Hamiltonian. Here, $\hat{V}_{AB}$
denotes  the direct interaction between dipole $A$ and dipole $B$,
while $\hat V_{AB'}$ and $\hat V_{BA'}$ denote the indirect
interaction between each dipole and the image induced by the other one
in the plate. According to this approach, the vdW potential $U_{AB}$
can be identified with the second-order energy shift
\begin{multline}
\label{deltaE}
\Delta E_{AB}=-\sum_{(n,m)\neq (0,0)}
\frac{\langle 0_{A}|\langle 0_{B}|\hat H_{\mathrm {int}}
|n_{A}\rangle|m_{B}\rangle}{\hbar(E_{A}^n+E_{B}^m
-E_{A}^0-E_{B}^0)}
\\
\times\,\langle n_{A}|\langle
m_{B}|\hat H_{\mathrm {int}}|0_{A}\rangle |
0_{B}\rangle.
\end{multline}
($E_{A(B)}^n$, $|n_{A(B)}\rangle$; atomic eigenenergies and
eigenstates, respectively). In this approach, $U_{AB}^{(0)}$
corresponds to the product of two direct interactions, so it is
negative in agreement with Eq.~(\ref{full}), because of the minus sign
on the r.h.s.\ of Eq.~(\ref{deltaE}). Accordingly, $U_{AB}^{(2)}$ is
due to the product of two indirect interactions and is also
negative---in agreement with Eq.~(\ref{full}). The terms containing
one direct and one indirect interaction are contained in
$U^{(1)}_{AB}$ and determine its sign. We can hence predict the sign
of $U^{(1)}_{AB}$ from a graphical construction of the image charges,
as sketched in Figs.~\ref{con-par}--\ref{mag-ver}.

Figure \ref{con-par} shows two electric dipoles in front of a
perfectly conducting plate in the parallel case. The configuration of
dipoles and images indicates repulsion between dipole $A(B)$ and
dipole $B'(A')$, so $U_{AB}^{(1)}$ is positive, in agreement with
Tab.~\ref{sign}. On the contrary, in the vertical case from
Fig.~\ref{con-ver} attraction is indicated, i.e., negative
$U_{AB}^{(1)}$, which is also in agreement with Tab.~\ref{sign}.

The case of two electric dipoles in front of a perfectly permeable
plate can be treated by considering two magnetic dipoles in front of a
perfectly conducting plate, as the two situations are equivalent due
to the duality between electric and magnetic fields in the absence of
free charges or currents. {F}rom Figs.~\ref{mag-par} (parallel case)
and \ref{mag-ver} (vertical case) it is apparent that the interaction
between dipole $A(B)$ and dipole $B'(A')$ is attractive in the
parallel case and repulsive in the vertical case, again confirming the
sign of $U_{AB}^{(1)}$ as given in Tab.~\ref{sign}.

When the dipole--dipole separation in Fig.~\ref{mag-ver} is
sufficiently small compared with the dipole--surface separations, then
the direct interaction between the two dipoles is expected to be
stronger than their indirect interaction via the image dipoles. As a
result, $U_{AB}^{(1)}$ will be the dominant term in
$U_{AB}^{(1)}\!+\!U_{AB}^{(2)}$ and
$U_{AB}^{(1)}\!+\!U_{AB}^{(2)}$ becomes positive. However, when the
dipole--dipole separation exceeds the dipole--surface separations,
then the indirect interaction may become comparable to the direct one,
and $U_{AB}^{(2)}$ may be the dominant term, leading to negative
$U_{AB}^{(1)}\!+\!U_{AB}^{(2)}$. The image dipole model hence gives
also a qualitative explanation of the condition~(\ref{condition}).
%%%%%%%%%%%%%%%%%%%%%%%%%%%%%%%% image parallel conductor%%%%%%%%%%%%%
%%%%%%%%%%%%%%%%%%%%%%%%%%%%%%%%%%%%%%%%%%%%%%%%%%%%%%%%%%%%%%%%%%%%%%
%%%%%%%%%%%%%%%%%%%%%%%%%%%%%%%% images vertical conductor %%%%%%%%%%%
%%%%%%%%%%%%%%%%%%%%%%%%%%%%%%%%%%%%%%%%%%%%%%%%%%%%%%%%%%%%%%%%%%%%%%
%%%%%%%%%%%%%%%%%%%%%%%%%%%%%%%%% images magnetic parallel%%%%%%%%%%%%
%%%%%%%%%%%%%%%%%%%%%%%%%%%%%%%%%%%%%%%%%%%%%%%%%%%%%%%%%%%%%%%%%%%%%%
%%%%%%%%%%%%%%%%%%%%%%%%%%%%%%%% images magnetic vertical%%%%%%%%%%%%%
%%%%%%%%%%%%%%%%%%%%%%%%%%%%%%%%%%%%%%%%%%%%%%%%%%%%%%%%%%%%%%%%%%%%%%

%%%%%%%%%%%%%%%%%%%%%%%%%%%%%%%%%%%%%%%%%%%%%%%%%%%%%%%%%%%%%%%%%%%%%%

\subsubsection{Magneto-electric half space}
\label{Sec3.2.2}

Let us now abandon the assumption of perfect reflectivity and 
consider a magneto-electric half space of permittivity
$\varepsilon(\omega)$ and permeability $\mu(\omega)$. In this case,
the reflection coefficients in Eqs.~(\ref{81}) and (\ref{82}) are
given by
\begin{align}
\label{eq95}
&r_s=r_s(q,u)=\frac{\mu(iu)b-b_\mathrm{M}}
 {\mu(iu)b+b_\mathrm{M}}\,,\\
\label{eq95b}
&r_p=r_p(q,u)=\frac{\varepsilon(iu)b\!-\!b_\mathrm{M}}
 {\varepsilon(iu)b+b_\mathrm{M}}\,,
\end{align}
where $b$ is defined by Eq.~(\ref{bl}), and
\begin{equation}
\label{bl2}
b_\mathrm{M}=b_\mathrm{M}(q,u)
=\sqrt{\varepsilon(iu)\mu(iu)\frac{u^2}{c^2}+q^2}\,.
\end{equation}

In the retarded limit, $l,z_A,z_B$ $\!\gg$ $\!c/\omega_{\mathrm
{min}}$ (where $\omega_\mathrm{min}$ now denotes the minimum of all
resonance frequencies of atoms $A$ and $B$ and the magneto-electric
medium) we may again approximate the atomic polarizabilities by their
static values, recall Eq.~(\ref{alpha0}), and similarly we may set
\begin{equation}
\label{Eq66}
\varepsilon(iu)\simeq\varepsilon(0),\qquad
 \mu(iu)\simeq\mu(0).
\end{equation}
Replacing the integration variable $q$ in Eq.~(\ref{81}) by 
\mbox{$v$ $\!=$ $\!bc/u$} [cf.\ Eq.~(\ref{C8})], 
one can show that the contribution $U_{AB}^{(1)}$ to the 
vdW potential takes the form
\begin{align}
&U_{AB}^{(1)}(\mathbf{r}_A,\mathbf{r}_B)
=\frac{\hbar c}{32\pi^3 l^3\varepsilon_0^2}
\alpha_{A}(0)\alpha_{B}(0)\int_1^\infty
{\rm d}v\,
\nonumber\\&\quad\times
\bigg(\bigg\{v^2\bigg[Z^2A_{5-}+(Z^2-2X^2)\bigg(\frac{A_{4-}}{l}
+\frac{A_{3-}}{l^2}\bigg)
\nonumber\\&\qquad
+l^2A_{5+}+lA_{4+}+A_{3+}\bigg]+2(v^2-1)\bigg[X^2B_{5}
\nonumber\\&\qquad
+\big(X^2-2Z^2\big)\bigg(\frac{B_4}{l}+\frac{B_3}{l^2}\bigg)\bigg]
\bigg\}r_{p}
\nonumber\\
&\qquad
+\bigg[Z^2A_{5+}+\big(Z^2-2X^2\big)\bigg(\frac{A_{4+}}{l}
+\frac{A_{3+}}{l^2}\bigg)
\nonumber\\&\qquad
+l^2A_{5-}+lA_{4-}+A_{3-}\bigg]r_{s}
\bigg),
\end{align}
where according to Eqs.~(\ref{eq95}) and (\ref{eq95b}), the static
reflection coefficients are given by
\begin{align}
\label{static1}
&r_s=r_s(v)
=\frac{\mu(0)v-\sqrt{\varepsilon(0)\mu(0)-1+v^2}}
{\mu(0)v+\sqrt{\varepsilon(0)\mu(0)-1+v^2}}\,, \\
\label{static2}
&r_p=r_p(v)
=\frac{\varepsilon(0)v-\sqrt{\varepsilon(0)\mu(0)-1+v^2}}
{\varepsilon(0)v+\sqrt{\varepsilon(0)\mu(0)-1+v^2}}\,,
\end{align}
and
\begin{align}
\label{A}
&A_{k\pm}=
 \int_0^\infty\mathrm d
 x\,x^k\,e^{-\lambda x}\big[J_0(\zeta x)\pm J_2(\zeta x)\big],\\
\label{B}
&B_k=\int_0^\infty{\mathrm d}x\,x^k
 e^{-\lambda x}J_0(\zeta x),
\end{align}
with  $\lambda$ $\!=$ $\!l$ $\!+$ $\!vZ_+$ and $\zeta$ $\!=$
$\!X\sqrt{v^2-1}$ (for explicit expressions of $A_{k\pm}$ and $B_k$,
see App.~\ref{an}). Similarly, Eq.~(\ref{82}) reduces to
\begin{align}
\label{24}
&U_{AB}^{(2)}=-\frac{\hbar c}{64\pi^3\varepsilon_0^2}\,\alpha_{A}(0)
\alpha_{B}(0)\int_1^\infty {\rm d}v\int_1^\infty {\rm d}v'
\nonumber\\&\quad\times
\biggl\{
\Big(r_{p}r_p'
\big[3v^2v'^2-2(v^2+v'^2)+2\big]
+r_{s}r_s'-r_{s}r_p'v'^2
\nonumber\\&\quad
-r_pr_s'v^2\Big)M_0
+4vv'\sqrt{v^2-1}\sqrt{v'^2-1}r_{p}r_p'M_1
\nonumber\\&\quad
+\big(r_{s}r_s'+r_{p}r_p'v^2v'^2+r_{s}r_p'v'^2
+r_pr_s'v^2\big)M_2\biggr\}
\end{align}
[$r'_{\sigma}=r_{\sigma}(v')$ for $\sigma$ $\!=$ $\!s,p$], where
\begin{equation}
\label{M}
M_\nu=\int_0^\infty{\mathrm d}x\,x^6e^{-(v+v')Z_+ x}J_\nu(\zeta
x)J_\nu(\zeta'x)
\end{equation}
($\zeta'$ $\!=$ $\!X\sqrt{v'^2-1}$), which can be evaluated
analytically only in some special cases. In particular, when $X$
$\!\ll$ $\! Z_+$, then approximately
\begin{equation}
M_\nu=
J_\nu^2(0)\int_0^\infty{\mathrm d}x\,
x^6e^{-(v+v')Z_+x}
= \frac{720\delta_{\nu0}}{(v+v')^7Z_+^7}\,.
\end{equation}

Analytic expressions for $U_{AB}^{(1)}$ and $U_{AB}^{(2)}$
in the nonretarded limit, $l,z_{A},z_{B}$ $\!\ll$
$\!c/[\sqrt{\varepsilon(0)\mu(0)}\,\omega_\mathrm{max}]$ [with
$\omega_\mathrm{max}$ being the maximum of all resonance frequencies
of atoms $A$ and $B$ and the magneto-electric medium], can be
obtained by using in Eqs.~(\ref{81}) and (\ref{82}), respectively, the
relevant elements of the scattering part of Green tensor as given in
App.~\ref{mulg}. In the case of a purely electric half space ($\mu$
$\!\equiv$ $\!1$) we derive [Eqs.~(\ref{Axx})--(\ref{Bzz})]
\begin{align}
\label{udie}
U_{AB}=&-\frac{C_6}{l^6}
+\frac{\big[4X^4-2Z^2Z_+^2+
X^2(Z^2+Z_+^2)\big]
D}{l^5l_+^5}\nonumber\\
&-\frac{E}{l_+^6}\,,
\end{align}
where $C_6$ is given by Eq.~(\ref{cnr}), and
\begin{align}
\label{c1}
&D=\frac{\hbar}{16\pi^3\varepsilon_0^2}
\int_0^\infty{\mathrm d}u\,\alpha_{A}(iu)
\alpha_{\mathrm
B}(iu)\frac{\varepsilon(iu)-1}{\varepsilon(iu)+1}\,,
\\
\label{c2}
&E=\frac{3\hbar}{16\pi^3\varepsilon_0^2}\int_0^\infty{\rm
d}u\,\alpha_{A}(iu)\alpha_{B}(iu)
\bigg[\frac{\varepsilon(iu)-1}{\varepsilon(iu)+1}
\bigg]^2.
\end{align}
In particular, in the limiting case when $l$ $\!\ll$ $\!Z_+$,
Eq.~(\ref{udie}) reduces to
\begin{equation}
\label{udie2}
U_{AB}=-\frac{C_6}{l^6}
+\frac{\big(X^2-2Z^2\big)D}{l^5Z_+^3}\,.
\end{equation}
It is seen that the second term on the r.h.s.\ of this equation is
positive  (negative) in the parallel (vertical) case, so the vdW
potential is reduced (enhanced) by the presence of the electric half
space.

In the case of a purely magnetic half space ($\varepsilon$
$\!\equiv$ $\!1$) we derive [Eqs.~(\ref{Axx2})--(\ref{Azz})]
\begin{equation}
\label{umag1}
U_{AB}=
-\frac{C_6}{l^6} +
\frac{\big[Z^2-2X^2+3Z_+(l_+-Z_+)\big]
 F}{l^5l_+}\,,
\end{equation}
where
\begin{multline}
F=\frac{\hbar}{64\pi^3\varepsilon_0^2}\int_0^\infty
{\rm d}u\,\Bigl(\frac{u}{c}\Bigr)^2\alpha_{A}(iu)\alpha_{B}(iu)\\
\times\;\frac{[\mu(iu)-1][\mu(iu)-3]}{\mu(iu)+1}\,.
\end{multline}
Note that $U_{AB}^{(2)}$ does not contribute to the asymptotic
nonretarded vdW potential $U_{AB}$ for the purely magnetic
half space. In particular in the limiting case when $X$ $\!\ll$
$\!Z_+$, Eq.~(\ref{umag1}) reduces to
\begin{equation}
\label{umag2}
U_{AB}=
-\frac{C_6}{l^6} +
\frac{\big(2Z^2-X^2\big)F}{2l^5Z_+}\,.
\end{equation}
It is seen that the second term in the r.h.s.\ of this equation is
negative (positive) in the  parallel (vertical) case, so the vdW
potential is enhanced (reduced) due to the presence of the magnetic
half space.

It should be pointed out that the nonretarded limit for the
magneto-electric half space is in general incompatible with the limit
of perfect reflectivity [\mbox{$\varepsilon(iu)\rightarrow \infty$} or
\mbox{$\mu(iu)$ $\!\to$ $\!\infty$}] considered in
Sec.~\ref{Sec3.2.1}, as is clearly seen from the condition given above
Eq.~(\ref{udie}) [cf.\ also the expansions (\ref{rs-lim}) and
(\ref{rp-lim}), which are not well-behaved in the limit of perfect
reflectivity]. As a consequence, Eq.~(\ref{umag1}) does not reduce to
Eq.~(\ref{full}) via the limit \mbox{$\mu(iu)$ $\!\to$ $\!\infty$}. It
is therefore remarkable that the result for a purely electric half
space, Eq.~(\ref{udie}), does reduce to Eq.~(\ref{full}) in the limit
\mbox{$\varepsilon(iu)$ $\!\to$ $\!\infty$}, as already noted in
Ref.~\cite{Babiker76} in the case of the single-atom potential.

%%%%%%%%%%%%%%%%%%%%%%%%%% POTENTIAL PARALLEL %%%%%%%%%%%%%%%%%%%%%
\begin{figure}
\noindent
\begin{center}
\includegraphics[width=\linewidth]{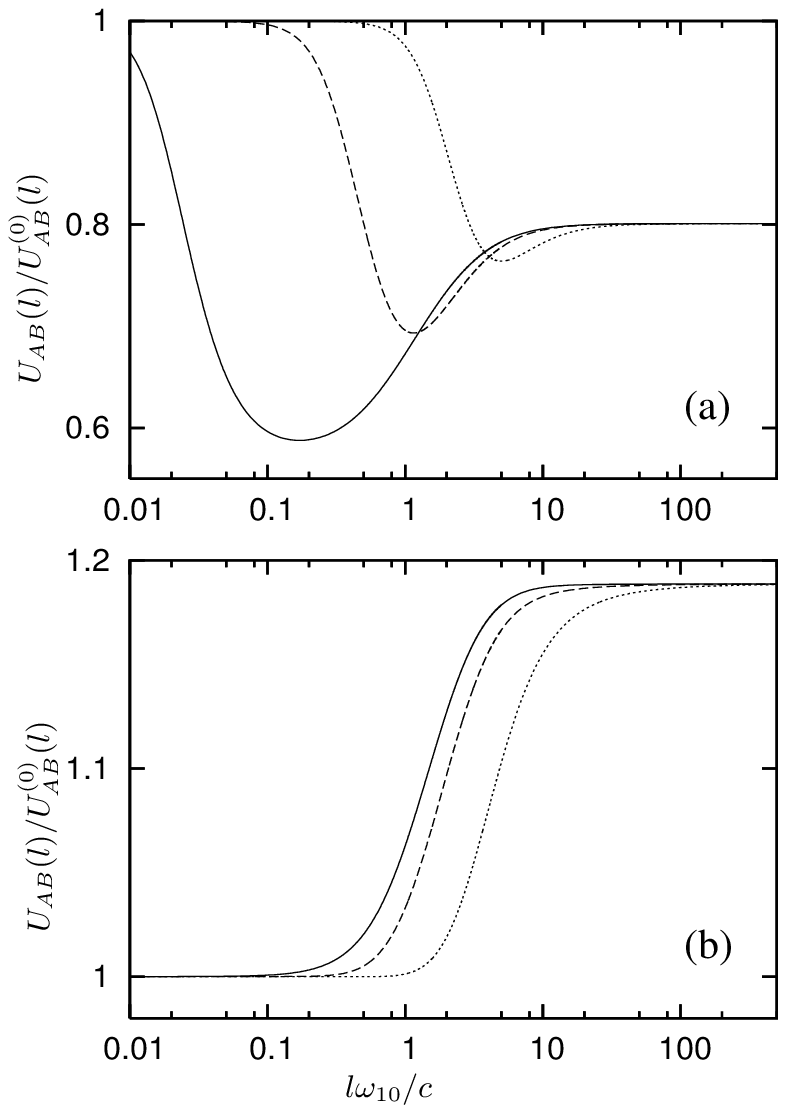}
\end{center}
\caption{
\label{p-par}
The normalized vdW potential of two atoms in the parallel case in
presence of (a) a  purely dielectric half space with
\mbox{$\omega_{\mathrm{P}e}/\omega_{10}$ $\!=$ $\!3$},
\mbox{$\omega_{\mathrm{T}e}/\omega_{10}$ $\!=$ $\!1$}, and
\mbox{$\gamma_e/\omega_{10}$ $\!=$ $\!0.001$} (b) a purely magnetic
half space with \mbox{$\omega_{\mathrm{P}m}/\omega_{10}$ $\!=$ $\!3$},
\mbox{$\omega_{\mathrm{T}m}/\omega_{10}$ $\!=$ $\!1$},
and \mbox{$\gamma_m/\omega_{10}=0.001$} is shown as a function of the
atom-atom separation $l$, with the atoms placed at at distance
$z_{A}\!=\!z_{B}$ $\!=$ $\!0.01c/\omega_{10}$ (solid line),
$0.2c/\omega_{10}$ (dashed line), and $c/\omega_{10}$ (dotted line)
from the half space.
}
\end{figure}%
%%%%%%%%%%%%%%%%%%%%%%%%%%%%%%%%%%%%%%%%%%%%%%%%%%%%%%%%%%%%%%%%%%%%%%
%%%%%%%%%%%%%%%%%%%%%%%%%%%% POTENTIAL VERTICAL %%%%%%%%%%%%%%%%%%%%%
\begin{figure}
\noindent
\begin{center}
\includegraphics[width=\linewidth]{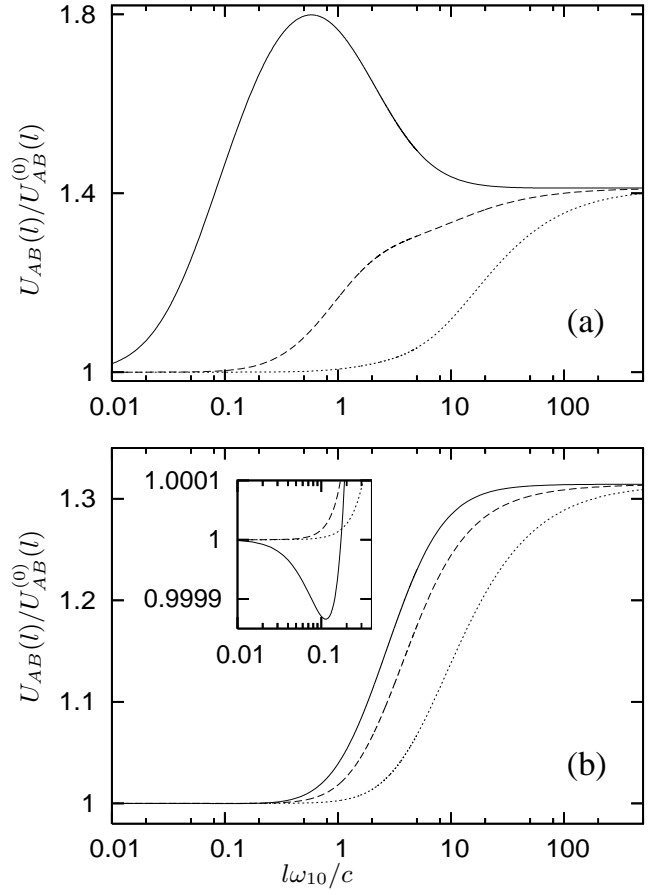}
\end{center}
\caption{
\label{p-ver}
The normalized vdW potential of two atoms in the vertical case in the
presence of (a) a purely dielectric half space and (b) a purely
magnetic half space is shown as a function of the atom-atom
separation $l$. The distance between atom $A$ (which is closer to the
surface of the half space than atom $B$) and the surface is $z_A$
$\!=$ $0.01c/\omega_{10}$ (solid line), $0.2c/\omega_{10}$ (dashed
line), and $c/\omega_{10}$ (dotted line). All other parameters are the
same as in Fig.~\ref{p-par}.}
\end{figure}%
%%%%%%%%%%%%%%%%%%%%%%%%% FORCE VERTICAL A %%%%%%%%%%%%%%%%%%%%%%%%%%
\begin{figure}
\noindent
\begin{center}
\includegraphics[width=\linewidth]{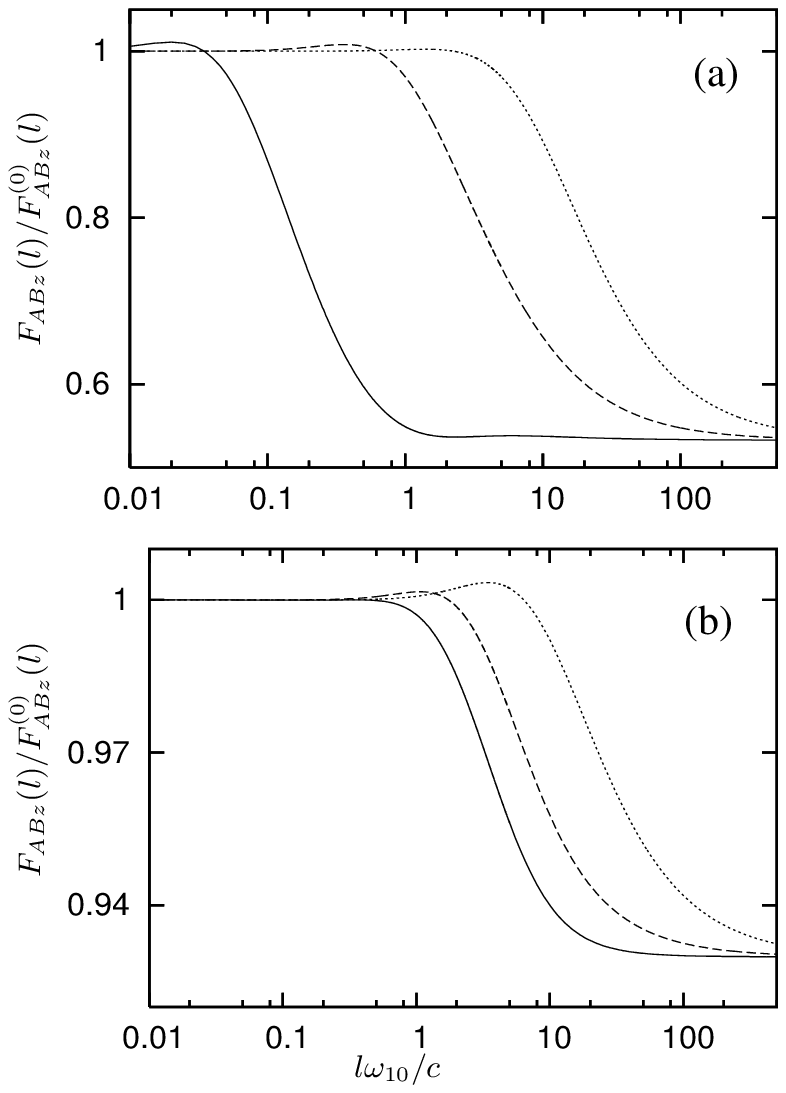}
\end{center}
\caption{
\label{f-a-ver}
The normalized vdW force acting on atom $A$ (which is closer to the
surface of the half space than atom $B$) in the presence of (a) a
purely dielectric half space and (b) a purely magnetic half space is
shown as a function of the atom-atom separation $l$. All parameters
are the same as in Fig.~\ref{p-ver}.}
\end{figure}%
%%%%%%%%%%%%%%%%%%%%%%%%%%%%%%%%%%%%%%%%%%%%%%%%%%%%%%%%%%%%%%%%%%%%%%
Figures~\ref{p-par}--\ref{f-a-ver} show the results of an exact
(numerical) calculation of the vdW interaction between two identical
atoms near a semi-infinite half space, as given by Eqs.~(\ref{uparts})
together with Eqs.~(\ref{Eq19}), (\ref{81}), and (\ref{82}) as well as
Eqs.~(\ref{eq95}) and (\ref{eq95b}). In the calculations, we have
used single-resonance models for both the polarizability of the atoms,
\begin{equation}
\label{alpha}
\alpha_A(\omega)=\alpha_B(\omega)
=\lim_{\epsilon\to 0}
 \frac{2}{3\hbar}
 \frac{\omega_{10}|\mathbf{d}_{10}|^2}{\omega_{10}^2-\omega^2
 - i\omega\epsilon}
\end{equation}
(with $\omega_{10}$ and $\mathbf{d}_{10}$ denoting the frequency and
electric dipole matrix element of the dominant atomic transition,
respectively), and the permittivity and permeability of the half
space,
\begin{equation}
\label{e109}
\varepsilon(\omega)=1+\frac{\omega_{\mathrm{P}e}^2}{\omega^2_{{\mathrm
{T}e}}
-\omega^2-i\omega\gamma_e}\,,
\end{equation}
\begin{equation}
\label{e110}
\mu(\omega)=1+\frac{\omega_{\mathrm{P}m}^2}{\omega^2_{\mathrm{T}m}
-\omega^2-i\omega\gamma_m}\,.
\end{equation}
In the figures the potentials and the forces are normalized w.r.t.\
their values in free space as given by Eq.~(\ref{Eq19}), so one can
clearly see that the vdW interaction is unaffected by the presence of
the half space for atom--half-space separations that are much greater
than the interatomic separations (the curves approaching unity for
$z_A,z_B$ $\!\gg$ $\!l$), while an asymptotic enhancement or reduction
of the interaction is observed in the opposite limit.

Figure~\ref{p-par}(a) shows the dependence of the normalized vdW
potential on the atom--atom separation $l$ in the parallel case
\mbox{($Z$ $\!=$ $\!0$)} for different values of the distance
\mbox{$z_A$ $\!=$ $\!z_B$} of the atoms from a purely dielectric
half space. The ratio of the interatomic force along the connecting
line of the two atoms, $F_{ABx}(l)$ [Eq.~(\ref{inforce})], to the
corresponding force in free space, $F^{(0)}_{ABx}(l)$, follows
closely the ratio $U_{AB}(l)/U^{(0)}_{AB}(l)$, so that, within the
resolution of the figures, the curves for
$F_{ABx}(l)/F^{(0)}_{ABx}(l)$ (not shown) would almost coincide with
those for $U_{AB}(l)/U^{(0)}_{AB}(l)$. The figure reveals that due to
the presence of the dielectric half space the attractive vdW potential
and force are reduced, in agreement with the predictions from the
nonretarded limit, Eq.~(\ref{udie2}). The relative reductions of the
potential and the force are not monotonic, there is a value of the
atom--atom separation where the reduction is strongest.

The $l$-dependence  of $U_{AB}(l)/U^{(0)}_{AB}(l)$ in the presence of
a purely magnetic half space in the parallel case is shown in
Figs.~\ref{p-par}(b). The corresponding force ratio
$F_{ABx}(l)/F^{(0)}_{ABx}(l)$ (not shown) again behaves like
$U_{AB}(l)/U^{(0)}_{AB}(l)$. The figure indicates
that the presence of a purely magnetic half space enhances the vdW
interaction between the two atoms, with the enhancement increasing
with the atom-atom separation, in agreement with the nonretarded
limit, Eq.~(\ref{umag2}).

Figure~\ref{p-ver} shows $U_{AB}(l)/U^{(0)}_{AB}(l)$ in the vertical
case \mbox{($X$ $\!=$ $\!0$)} when the half space is purely dielectric
[Fig.~\ref{p-ver}(a)] or purely magnetic [Fig.~\ref{p-ver}(b)]. In the
figures, atom $A$ is assumed to be closer to the surface of the half
space than atom $B$, and the graphs show the variation of the vdW
potential with the atom-atom separation $l$ for different distances
$z_A$ of atom $A$ from the half space. It is seen that for a purely
dielectric half space the potential is enhanced compared to the one
observed in the free-space case---in agreement with Eq.~(\ref{udie2}).
Note that there are values of the atom--atom separation at which the
enhancement is strongest.

For a purely magnetic half space, the potential is seen to be
typically enhanced although for very small atom--atom separations a
reduction appears [inset in Fig.~\ref{p-ver}(b)]---in agreement with
Eq.~(\ref{umag2}). Due to this slight reduction for small atom--atom
separations, the relative enhancement is not monotonous, in contrast
to what is suggested by the large figure.

Whereas the force $F_{BAz}(l)/F^{(0)}_{BAz}(l)$ for the
force acting on atom $B$ (not shown) again follows closely the
potential ratio $U_{AB}(l)/U^{(0)}_{AB}(l)$ for both dielectric and
magnetic half spaces (as in Fig.~\ref{p-ver}, not shown), the ratio
$F_{ABz}(l)/F^{(0)}_{ABz}(l)$, for the force acting on atom $A$
noticeably differs from $U_{AB}(l)/U^{(0)}_{AB}(l)$
(Fig.~\ref{f-a-ver}). Clearly, the difference is due to the fact that
the atom $A(B)$ which is responsible for the force $F_{BA(AB)z}(l)$ is
situated on the same side of atom $B(A)$ as the half space in the
former case, but on a different side in the latter case
(cf.~Figs.~\ref{con-ver} and \ref{mag-ver}).

%%%%%%%%%%%%%%%%%%%%%%%%%%%%%%%%%%%%%%%%%%%%%%%%%%%%%%%%%%%%%%%%%%%%%%

\section{Summary and Conclusions}
\label{Sec4}

Starting from the CP potential of a single polarizable atom placed
within a given arrangement of magneto-electric bodies, we have
presented a macroscopic derivation of two-atom vdW potentials: By
introducing an additional weakly polarizable body and linearly
expanding the resulting CP interaction in terms of the body's
susceptibility, the vdW potential between two polarizable atoms in the
presence of an arbitrary arrangement of dispersing and absorbing
magneto-electric bodies has been inferred. The vdW potential between
a polarizable atom and a magnetizable one has been derived in a
similar way by introducing a weakly magnetizable body. The general
formulas have been used to study the influence of polarizability and
magnetizability on the vdW potential between two atoms in free space.
In particular, it has been shown that the vdW interaction of a
polarizable atom with a magnetizable one is always repulsive, in
contrast to the well-known attractive potential between two
polarizable atoms.

To illustrate the influence of the presence of magnetodielectric
bodies on the vdW potential, we have considered the example of two
polarizable atoms near a perfectly reflecting plate. It has turned out
that due to the presence of the plate the attractive vdW interaction
between the atoms can be enhanced or reduced depending on the
magneto-electric properties of the plate and the specific alignment
of the atoms with respect to the plate. In particular, in the
nonretarded limit these effects can be qualitatively explained using
the method of image dipoles. To be more realistic, we have also
calculated the vdW potential for the case of the two atoms near a
magneto-electric half space of finite permittivity and permeability.
The analytical results show that in the nonretarded limit the
potential in the case of a purely electric half space is reduced
(enhanced) compared to its value in free space in the case of parallel
(vertical) alignment of the two atoms, while in the case of a purely
magnetic half space it is enhanced (reduced) for parallel (vertical)
alignment of the two atoms. The numerical computation of the potential
in the whole distance regime confirms the analytical results. In
addition, it shows that the relative enhancement/reduction of the vdW
interaction is not always monotonous, but may in general display
maxima or minima, in particular in the case of a purely dielectric
half space.

In conclusion, the examples studied in this work suggest that the sign
of the vdW potential is entirely determined by the electric/magnetic
nature of the interacting atoms, while the strength of the respective
attractive or repulsive potentials can be controlled by the presence
of magneto-electric bodies.

%%%%%%%%%%%%%%%%%%%%%%%%%%%%%%%%%%%%%%%%%%%%%%%%%%%%%%%%%%%%%%%%%%%%%%

\begin{acknowledgments}
This work was supported by the Deutsche Forschungsgemeinschaft.
H.S.\ would like to thank the Ministry of Science, Research, and
Technology of Iran. H.T.D.\ is grateful to T.\ Kampf for a helpful
hint on programming. He would also like to thank the Alexander von
Humboldt Stiftung and the National Program for Basic Research of
Vietnam.
\end{acknowledgments}

%%%%%%%%%%%%%%%%%%%%%%%%%%%%%%%%%%%%%%%%%%%%%%%%%%%%%%%%%%%%%%%%%%%%%%

\begin{appendix}

%%%%%%%%%%%%%%%%%%%%%%%%%%%%%%%%%%%%%%%%%%%%%%%%%%%%%%%%%%%%%%%%%%%%%%

\section{Scattering Green tensor for a semi-infinite half space}
\label{mulg}

The scattering Green tensor for a semi-infinite half space can be
given in the form \cite{chew}
\begin{equation}
\label{mul-a}
\bm{G}^{(1)}(\mathbf{r},\mathbf{r}',iu) =
\int\mathrm{d}^2q\,
e^{i\mathbf{q}\cdot(\mathbf{r}-\mathbf{r}')}
\bm{G}^{(1)}(\mathbf{q},z,z',iu)
\end{equation}
($\mathbf{q}\perp\mathbf{e}_z$), where
\begin{equation}
\label{mul-b}
\bm{G}^{(1)}(\mathbf{q},z,z',iu) = \frac{1}{8\pi^2b}
 \sum_{\sigma=s,p}\mathbf{e}_\sigma^+\mathbf{e}_\sigma^- r_\sigma
 e^{-b(z+z')},
 \end{equation}
with
\begin{align}
\label{es}
&\mathbf{e}_s^\pm=\sin{\phi}\,{\mathbf
e}_x-\cos\phi\,{\mathbf e}_y,
\\
\label{ep}
&\mathbf{e}_p^\pm=\mp\frac{b}{k}(\cos\phi\,{\mathbf e}_x+
\sin\phi\,{\mathbf e}_y)-\frac{iq}{k}\,{\mathbf e}_z
\end{align}
($\mathbf{e}_q$ $\!=$ $\!\cos\phi\,\mathbf{e}_x\!+
\!\sin\phi\,\mathbf{e}_y$ $\!=$ $\!\mathbf{q}/q$, $q$ $\!=$
$\!|\mathbf{q}|$) denoting the polarization vectors for $s$- and
$p$-polarized waves propagating in the positive($+$)/negative($-$)
$z$-direction. Further, $b$ and $k$ are defined according to
Eqs.~(\ref{bl}) and (\ref{kl}), respectively, and the reflection
coefficients $r_s,r_p$ are given by Eq.~(\ref{rp-ref}) for a
perfectly reflecting plate and by Eqs.~(\ref{eq95}) and (\ref{eq95b})
for a magneto-electric half space. Equations~(\ref{es}) and
(\ref{ep}) imply that
\begin{equation}
\mathbf e_{s}^+\mathbf e_{s}^{-}=
\left(
\begin{array}{lll}
\sin^2\phi & -\sin\phi\cos\phi & 0 \\
-\sin\phi\cos\phi & \cos^2\phi & 0 \\
0 & 0 & 0
\end{array}
\right),
\end{equation}
\begin{eqnarray}
&&\mathbf e_{p}^+\mathbf e_{p}^{-}=\nonumber\\[1ex]
&&\!\left(
\begin{array}{lll}
-\frac{b^2}{k^2}\cos^2\phi & -\frac{b^2}{k^2}
\sin\phi\cos\phi & \frac{ibq}{k^2}\cos\phi \\
-\frac{b^2}{k^2}\sin\phi\cos\phi &-\frac{b^2}{k^2}
\sin^2\phi &\frac{ibq}{k^2}\sin\phi \\
-\frac{ibq}{k^2}\cos\phi & -\frac{ibq}{k^2}\sin\phi  &
-\frac{q^2}{k^2}
\end{array}
\right).\nonumber\\
\end{eqnarray}
Substituting these results into Eqs.~(\ref{mul-a}) and (\ref{mul-b}),
performing the $\phi$-integrals by means of~\cite{abra}
\begin{equation}
\int_0^{2\pi}{\mathrm d}\phi\,e^{ix\cos{\phi}}\cos(\nu x)=2\pi
i^{\nu} J_{\nu}(x),
\end{equation}
and using the relation
\begin{equation}
\frac{J_1(x)}{x}=\frac{J_0(x)-J_2(x)}{2}\,,
\end{equation}
we arrive at the Eqs.~(\ref{62})--(\ref{64}).

In the particular case of a perfectly reflecting plate in the retarded
limit, it is convenient to replace the integration variable $q$
in Eqs.~(\ref{62})--(\ref{64}) in favor of \mbox{$v$ $\!=$
$\!bc/u$}, i.e., $q$ $\!=$ $\!\sqrt{v^2-1}\,u/c$ [see Eq.~(\ref{bl})],
and hence
\begin{equation}
\label{C8}
\int_0^\infty{\mathrm d}q\,\frac{q}{b}\cdots
\ \mapsto\ \int_1^\infty
 {\mathrm d}v\,\frac{u}{c}\cdots\,.
\end{equation}
For $X$ $\!\ll$ $\!Z_+$, the exponential terms effectively limits the
integrals in Eqs.~(\ref{62})--(\ref{64}) to the region where
\mbox{$qX$ $\!\ll$ $\!1$}, hence we can approximate \mbox{$J_\nu(qX)$
by $J_\nu(0)$ $\!=$ $\!\delta_{\nu 0}$}, such that the nonzero
scattering-Green tensor components read
\begin{align}
\label{rx}
&G^{(1)}_{xx}(\mathbf r_{A},\mathbf r_{B},iu)
 =G^{(1)}_{yy}(\mathbf r_{A},\mathbf r_{B},iu)\nonumber\\
&\quad =\frac{1}{8\pi Z_+}\bigg[r_s-\bigg(1+2\,\frac{c}{uZ_+}+
 2\,\frac{c^2}{u^2Z_+^2}\bigg)r_p\bigg]e^{-uZ_+/c},\\
\label{rz}
&G^{(1)}_{zz}(\mathbf r_{A},\mathbf r_{B},iu)\nonumber\\
&\quad =-\frac{1}{2\pi Z_+}\bigg(\frac{c}{uZ_+}+
 \frac{c^2}{u^2Z_+^2}\bigg)r_p\,e^{-uZ_+/c},
\end{align}
leading to Eqs.~(\ref{gl79}) and (\ref{gl80}), recall
Eq.~(\ref{alpha0}).

In the nonretarded limit it can be shown that the main contribution to
the frequency integrals comes from the region where $u/(cb)$ $\!\ll$
$\!1 $ (cf.\ Ref.~\cite{Thomas}). In this region we have
\begin{equation}
\label{9}
q=b\sqrt{1-\frac{u^2}{b^2c^2}}\simeq b.
\end{equation}
By changing the integration variable $q$ according to
\begin{equation}
\label{change}
 \int_0^\infty{\rm d}q\,\frac{q}{b}\,
 \ldots
\ \mapsto\
\int_{u/c}^\infty\mathrm{d}b\,
\ldots
\end{equation}
and setting the lower limit of integration to zero, from
Eqs.~(\ref{62})--(\ref{64}) we find, after some algebra, the nonzero
elements of the scattering Green tensor to be approximately given by
\begin{align}
\label{xx}
& G^{(1)}_{xx}(\mathbf r_{A},\mathbf r_{B},iu)=
\frac{2X^2-Z_+^2}{4\pi l_+^5}\,\frac{c^2r_p}{
u^2}\,,\\
\label{yy}
& G^{(1)}_{yy}(\mathbf r_{A},\mathbf r_{B},iu)=-
\frac{1}{4\pi l_+^3}\,\frac{c^2r_p}{
u^2}\,,\\
\label{xz}
& G^{(1)}_{xz(zx)}
(\mathbf r_{A},\mathbf
r_{B},iu)=\MP
\frac{3XZ_+}{4\pi l_+^5}\,\frac{c^2r_p}{ u^2 }\,,\\
\label{zz}
& G^{(1)}_{zz}(\mathbf r_{A},\mathbf r_{B},iu)=
\frac{X^2-2Z_+^2}{4\pi l_+^5}\,\frac{c^2r_p}{u^2}\,,
\end{align}
with $l_+$ $\!=$ $\!\sqrt{X^2+Z_+^2}$, leading to Eqs.~(\ref{e1}) and
(\ref{e2}).

For a magneto-electric half space in the nonretarded limit, we apply
a similar procedure as below Eq.~(\ref{rz}) and expand the reflection
coefficients given by Eqs.~(\ref{eq95}) and (\ref{eq95b}) in terms of
$u/(bc)$,
\begin{align}
\label{rs-lim}
&r_s\simeq\frac{\mu(iu)-1}{\mu(iu)+1}-\frac{\mu(iu)[
\varepsilon(iu)\mu(iu)-1]}{[\mu(iu)+1]^2}\frac{u^2}{b^2c^2}\,,
\\
\label{rp-lim}
&r_p\simeq\frac{\varepsilon(iu)-1}{\varepsilon(iu)+1}-\frac{
\varepsilon(iu)[\varepsilon(iu)\mu(iu)-1]}{[\varepsilon(iu)+1]^2}
\frac{u^2}{b^2c^2}\,.
\end{align}
Substituting (\ref{rs-lim}) and (\ref{rp-lim}) into
Eqs.~(\ref{62})--(\ref{64}) and keeping only the leading-order terms
of $u/bc$, in the case of the purely electric half space we can
ignore $r_s$ and the second term in the r.h.s.\ of Eq.~(\ref{rp-lim}),
so the relevant elements of the scattering Green tensor are
approximately
\begin{align}
\label{Axx}
 &G^{(1)}_{xx}(\mathbf r_{A},\mathbf r_{B},iu)
 =\frac{2X^2-Z_+^2}{4\pi l_+^5}\,
 \frac{c^2}{u^2}\,
 \frac{\varepsilon(iu)-1}{\varepsilon(iu)+1}\,,\\
\label{Byy}
&G^{(1)}_{yy}(\mathbf r_{A},\mathbf r_{B},iu)
 =-\frac{1}{4\pi l_+^3}\,
\frac{c^2}{u^2}\,\frac{\varepsilon(iu)-1}{\varepsilon(iu)+1}\,,\\
\label{Axz}
&G^{(1)}_{xz(zx)}(\mathbf r_{A},\mathbf r_{B},iu)
=\MP\frac{3XZ_+}{4\pi l_+^5}\,
\frac{c^2}{u^2}\,
\frac{\varepsilon(iu)-1}{\varepsilon(iu)+1}\,,\\
\label{Bzz}
&G^{(1)}_{zz}(\mathbf r_{A},\mathbf r_{B},iu)
 =\frac{X^2-2Z_+^2}{4\pi l_+^5}\,
\frac{c^2}{u^2}\,
\frac{\varepsilon(iu)-1}{\varepsilon(iu)+1}\,.
\end{align}
For a purely magnetic half space, the first term on the r.h.s.\ of
Eq.~(\ref{rp-lim}) vanishes, so the leading order of $u/bc$ is due to
the second term as well as the first term on the r.h.s.\ of
Eq.~(\ref{rs-lim}), so the nonzero elements of the scattering Green
tensor can be approximated by
\begin{align}
\label{Axx2}
 G^{(1)}_{xx}(\mathbf r_{A},\mathbf r_{B},iu)
 =&\;\frac{ l_+-Z_+}{4 \pi X^2}
\frac{\mu(iu)-1}{\mu(iu)+1}
\nonumber\\
&+\frac{Z_+l_+-Z_+^2}{16\pi X^2l_+}
[\mu(iu)-1],
\end{align}
\begin{align}
\label{Ayy}
 G^{(1)}_{yy}(\mathbf r_{A},\mathbf r_{B},iu)=&\;
 \frac{l_+-Z_+}{16 \pi  X^2}
 [\mu(iu)-1]
\nonumber\\
&+\frac{Z_+l_+-Z_+^2}{4 \pi X^2l_+}
\frac{\mu(iu)-1}{\mu(iu)+1}\,,
\end{align}\vspace*{-3ex}
\begin{align}
\label{Axz2}
G^{(1)}_{xz(zx)}(\mathbf r_{A},\mathbf r_{B},iu)
=&\;\PM\frac{l_+-Z_+}{16 \pi Xl_+}
 [\mu(iu)-1],\hspace{4.5ex}
\end{align}\vspace*{-3ex}
\begin{align}
\label{Azz}
 G^{(1)}_{zz}(\mathbf r_{A},\mathbf r_{B},iu)=\frac{1}{16 \pi l_+}
[\mu(iu)-1].\hspace{7ex}
\end{align}

%%%%%%%%%%%%%%%%%%%%%%%%%%%%%%%%%%%%%%%%%%%%%%%%%%%%%%%%%%%%%%%%%%%%%%

\section{Explicit forms of $\bm{A_{k\pm}}$ and $\bm{B_k}$ in
Eqs.~(\ref{A}) and (\ref{B})}
\label{an}

The integrals in Eqs.~(\ref{A}) and (\ref{B}) can be performed to
obtain the following explicit expressions:
\begin{align}
&A_{3+}=\frac{6\lambda}
 {\bigl(\lambda^2+\zeta^2\bigr)^\frac{5}{2}}\,,
\end{align}\vspace*{-3ex}
\begin{align}
&A_{3-}=\frac{6\bigl(\lambda^3-4\lambda\zeta^2\bigr)}
 {\bigl(\lambda^2+\zeta^2\bigr)^\frac{7}{2}}\,,
\end{align}\vspace*{-3ex}
\begin{align}
&A_{4+}=
\frac{6\bigl(4\lambda^2-\zeta^2\bigr)}
{\bigl(\lambda^2+\zeta^2\bigr)^\frac{7}{2}}\,,
\end{align}\vspace*{-3ex}
\begin{align}
&A_{4-}=
\frac{6\bigl(4\lambda^4-27\lambda^2\zeta^2+4\zeta^4\bigr)}
 {\bigl(\lambda^2+\zeta^2\bigr)^\frac{9}{2}}\,,
\end{align}\vspace*{-3ex}
\begin{align}
&A_{5+}=\frac{30\bigl(4\lambda^3-3\lambda\zeta^2\bigr)}
{\bigl(\lambda^2+\zeta^2\bigr)^\frac{9}{2}}\,,
\end{align}\vspace*{-3ex}
\begin{align}
& A_{5-}=
 \frac{30\bigl(4\lambda^5-41\lambda^3\zeta^2+18\lambda\zeta^4\bigr)}
 {\bigl(\lambda^2+\zeta^2\bigr)^\frac{11}{2}}\,,
\end{align}\vspace*{-3ex}
\begin{align}
&B_{3}=\frac{3\lambda\bigl(2\lambda^2-3\zeta^2\bigr)}
{\bigl(\lambda^2+\zeta^2\bigr)^\frac{7}{2}}\,,
\end{align}\vspace*{-3ex}
\begin{align}
&B_{4}=\frac{3\bigl(8\lambda^4-24\lambda^2\zeta^2+3\zeta^4\bigr)}
{\bigl(\lambda^2+\zeta^2\bigr)^\frac{9}{2}}\,,
\end{align}\vspace*{-3ex}
\begin{align}
&B_{5}=\frac{15\lambda\bigl(8\lambda^4
-40\lambda^2\zeta^2+15\zeta^4\bigr)}
{\bigl(\lambda^2+\zeta^2\bigr)^\frac{11}{2}}\,.
\end{align}

%%%%%%%%%%%%%%%%%%%%%%%%%%%%%%%%%%%%%%%%%%%%%%%%%%%%%%%%%%%%%%%%%%%%%%

\end{appendix}

%%%%%%%%%%%%%%%%%%%%%%%%%%%%%%%%%%%%%%%%%%%%%%%%%%%%%%%%%%%%%%%%%%%%%%

\end{document}